\documentclass[aps,prl,reprint,superscriptaddress]{revtex4-1}

\usepackage{graphicx}
\usepackage{caption}
\usepackage{subcaption}
\usepackage{color}
\usepackage{cancel}
\usepackage{amsmath}
\usepackage[linktocpage,colorlinks=true,linkcolor=blue,citecolor=blue,breaklinks=true]{hyperref}
\usepackage[normalem]{ulem}
\usepackage{comment}
\usepackage[export]{adjustbox}
\newcommand{\ju}[1]{\textcolor{magenta}{{#1}}}

\begin{document}

%\preprint{}

\title{Flow states and transitions of an active nematic in a three dimensional channel}
%\title{
%Route towards collective order in panicked human crowds
\author{Santhan Chandragiri}
 \affiliation{Department of Chemical Engineering, Indian Institute of Technology Madras, Chennai 600036, India}
\email[]{ch15d413@smail.iitm.ac.in}
%\thanks{}

\author{Amin Doostmohammadi}
\email{doostmohammadi@nbi.ku.dk}
\affiliation{Niels Bohr Institute, University of Copenhagen, Blegdamsvej 17, 2100 Copenhagen, Denmark}

\author{Julia M. Yeomans}
\email{julia.yeomans@physics.ox.ac.uk}
\affiliation{The Rudolf Peierls Centre for Theoretical Physics, Clarendon Laboratory, Parks Road, Oxford, OX1 3PU, UK.}

\author{Sumesh P. Thampi}
\email{sumesh@iitm.ac.in}
\affiliation{Department of Chemical Engineering, Indian Institute of Technology Madras, Chennai 600036, India}

%\homepage[]{http://www.tifrh.res.in/tcis/tcis-faculty/rama-govindarajan.html}

%\newcommand*{\ad}{\textcolor{blue}}
\date{\today} 

\begin{abstract}
We use active nematohydrodynamics to study the flow of an active fluid in a 3D microchannel, finding a transition between active turbulence and regimes where there is a net flow along the channel. We show that the net flow is only possible if the active nematic is flow aligning and that -- in agreement with experiments -- the appearance of the net flow depends on the aspect ratio of the channel cross-section. We explain our results in terms of  the when hydrodynamic screening due to the channel walls allows the emergence of vortex rolls across the channel.

%{Spontaneous flow generation due to collective dynamics is a generic feature of active fluids, making them a promising candidate for new self-sustained microfluidic applications. Despite theoretical, numerical, and experimental evidences of the emergence of coherent flows in two-dimensional microchannels, recent experiments have reported transition from turbulence to coherent flows in three-dimensional channels upon decreasing the aspect ratio of the channel cross section. This puzzling observation can not be explained based on the current understanding of 2D active fluids and indicates an intrinsically 3D phenomenon. In this article, we use simulations of active isotropic fluid to explore possible flow states and explain the underlying mechanism of transition to coherent flows in 3D confined active fluids. In addition to - in agreement with the experimental data - consolidating the role of channel aspect ratio, we also pinpoint the crucial role of the flow-aligning parameter (a material property) \ad{\sout{and the activity-induced nematic ordering}} in establishing a coherent flow, thus shedding light on the relevant regime where the experiments operate.}

\end{abstract}


%\keywords{}

\maketitle
% \section{Introduction}
% \textit{Introduction:-} 
Spontaneous flow generation associated with collective dynamics in cell colonies~\cite{Saw2017,Duclos2018}, bacterial suspensions~\cite{Wensink2012,Dunkel2013} and cytoskeletal elements, both {\it in vivo}~\cite{Kumar2014} and {\it in vitro}~\cite{Kruse2004,Schaller2010}, has been of great scientific interest in recent years~\cite{Koch2011,Ramaswamy2010,Marchetti2013,Gompper2020}. The continuous throughput of energy in these active systems manifests as active turbulence, a flow field characterised by spatio-temporally evolving fluid jets and vortices. 
Under the right conditions, such systems are capable of self-organization from chaotic flows into coherent flows: groups of active particles move together as a unit in a directed manner~\cite{Bricard13,Duclos2018,Wioland2016,Wu2017}. Coherent active flows have relevance to the
%However, the exact nature of these “right conditions” is currently the focus of intense study as many biological processes - including subcellular flows~\cite{Suzuki2017},
formation of bacterial biofilms~\cite{Conrad18}, wound healing~\cite{Poujade07}, organ formation~\cite{Mclennan12}, and collective tumor invasion~\cite{Clark15}. % – demonstrate active coherent flows. 
Beyond biological implications, understanding how these self-sustained flows can be controlled and directed would prove to be a tremendous advance in 
%the emerging field of
microfluidics~\cite{Whitesides06} where, conventionally, external flows are imposed for targeted drug delivery~\cite{Beebe}, for mixing in micro-reactors~\cite{Nguyen}, or for pumping fluids at microscales~\cite{Laser}. An outstanding question is therefore how the chaotic motion of active matter can be translated into self-sustained coherent flow.
%However, experiments show that confinement can significantly influence the flow states in active systems \cite{Duclos2018,Wioland2016,Wu2017}. Such studies are useful because they provide fundamental insights into the nonequilibrium physics that sustains  active turbulence. Moreover the ability to gain control of flow states by the confinement helps us in developing microfluidic devices for energy extraction \cite{Leonardo2010} and targeted drug delivery \cite{Gompper2020} applications. In this \textit{letter}, using active nematics as a model fluid, we demonstrate  how geometrical confinement, material properties and flow parameters can be simultaneously tuned to gain control over otherwise turbulent like flows in active systems.

Several lines of evidence demonstrate that geometric confinement can stabilize chaotic active motion into directed flows~\cite{Bricard13,Duclos2018,Wioland2016,Wu2017}. In two-dimensions theoretical analysis~\cite{Voituriez2005} predicts a transition from a quiescent fluid to a coherent, laminar flow through the channel, as the channel height increases beyond a well-defined threshold for a given activity. The properties of this transition are
%existence of such a transition and dependence on channel height is 
reproduced quantitatively in experiments on fibroblast cells confined in 2D channels of varying width~\cite{Duclos2018}. Beyond this linear instability, 2D experiments~\cite{Sanchez2012, Wioland2016, Opathalage2019, Hardouin2019, Suzuki2017}
and simulations~\cite{Shendruk2017, Chandragiri2019, Chen2018} have shown that oscillating flows and vortex-lattice states also emerge by increasing the channel height before transitioning to active turbulence.
%Experimentally, turbulent~\cite{Sanchez2012}, doubly periodic~\cite{Opathalage2019}, vortex-lattices~\cite{Hardoüin2019}, and unidirectional flows~\cite{Suzuki2017} have been observed in suspensions of microtubules and kinesin molecular motors residing on an oil-water interface. Similarly, successful comparisons between theory and experiments explaining the role of confinement in planar suspensions of \textit{Bacillus subtilis} \cite{Wioland2016} have been reported. 
 Despite this extensive research on 2D systems, the understanding of 3D active flows is in its infancy~\cite{Shendruk18,Binysh20,Duclos20}. Of particular interest, recent puzzling experiments show that the chaotic motion of a suspension of microtubules and molecular motors 
 can be stabilized into a coherent flow in meter-long channels with a square cross-section of any size, but not in channels with a rectangular cross-section~\cite{Wu2017}.
 (In this \textit{letter}, we will use the term coherent flows to refer to a state of broken symmetry along the channel length, whereby a net transport of fluid through the channel is obtained.)
 %  It was demonstrated that it is possible for a suspension of microtubules and molecular motors to autonomously flow through 3D, meter-long channels. 
 The maximum flow velocity reached about $10\,\text{mm}\,\text{s}^{-1}$, which is comparable to the velocities of pump-driven flows routinely used in microfluidics. However, the physical mechanism behind the transition to a net flow, and the reason for its dependence on the aspect ratio of the channel, remain unexplained. 
% why changing only the aspect ratio of channel cross-section results in transition between turbulence and coherent flows and its underlying physical mechanisms remain unexplained. 

Here, to address these outstanding questions and to bridge the gap between 2D and 3D studies of confined active flows, we perform numerical simulations of active nematohydrodynamics. 
Similarly to the experimental conditions \cite{Sanchez2012,Wu2017} the system is maintained above the isotropic-nematic transition temperature such that all the nematic ordering is activity-induced \ju{\cite{Santhosh2019}}.
We find that an active fluid transitions from active turbulence to coherent flows as the aspect ratio of the confining channel is reduced.
%in the absence of any thermodynamic ordering mechanism, an isotropic active fluid transitions between active turbulence and coherent flows as the aspect ratio of the 3D channel cross section is reduced - going from rectangular towards square cross sections.
In addition, the
simulations allow us to show that a necessary physical condition for such a crossover  is that the system must be
%transition to coherent flows is only possible in channels with small aspect ratios if the system is 
in the flow-aligning regime, where the induced order aligns with the self-generated shear flows. We explain the underlying mechanism of the transition to coherent flows based on these observations.\\

\begin{figure*}
\begin{subfigure}{0.3\linewidth}
    \includegraphics[width=\linewidth,trim={0 0 0 0},clip]{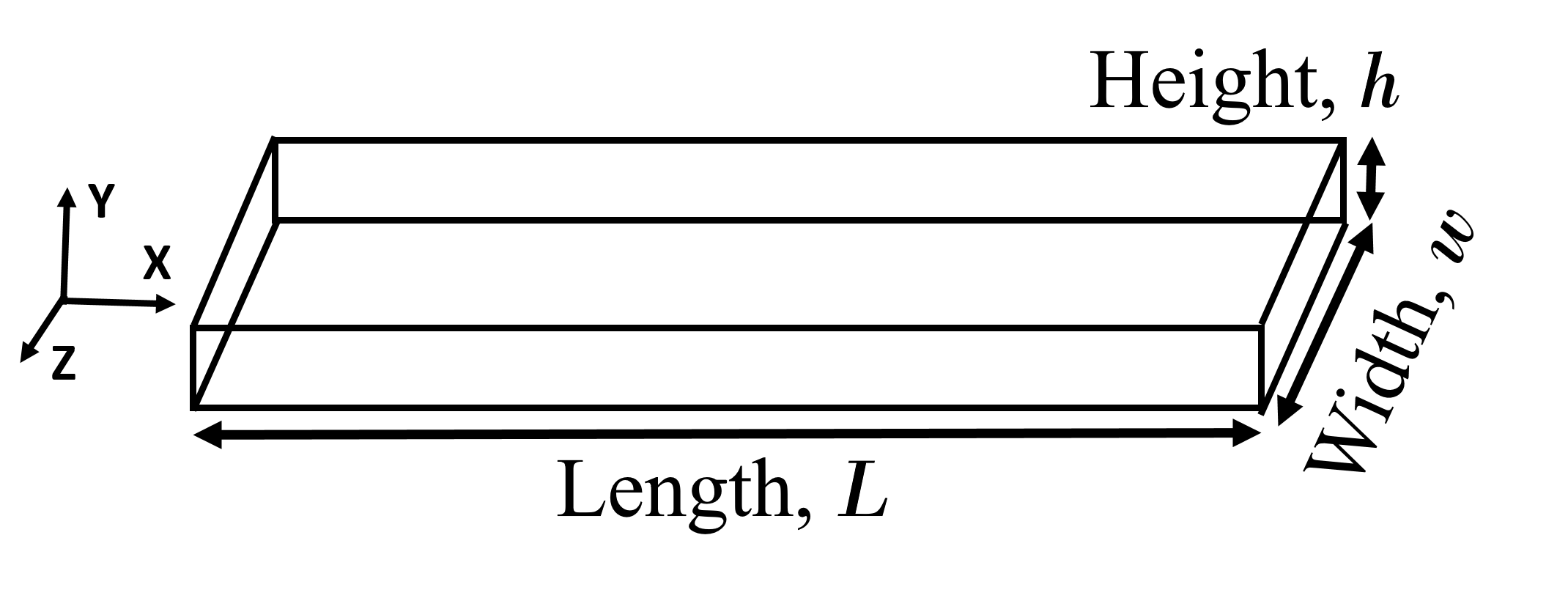} 
\caption{}
 \label{fig:schematic}
\end{subfigure}
\begin{subfigure}{0.3\linewidth}
    \includegraphics[width=1.7\linewidth,trim={8cm 0 -4cm 0},clip]{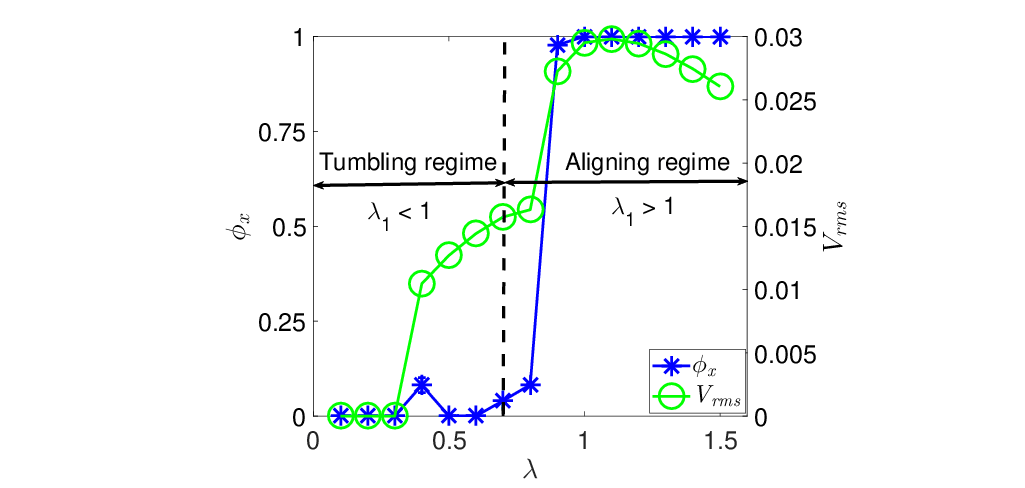} 
\caption{}
 \label{fig:lambda_sensitivity}
\end{subfigure}
\hspace{1cm}
\begin{subfigure}{0.3\linewidth}
    \includegraphics[width=1.3\linewidth,trim={8cm 0 4cm 0},clip]{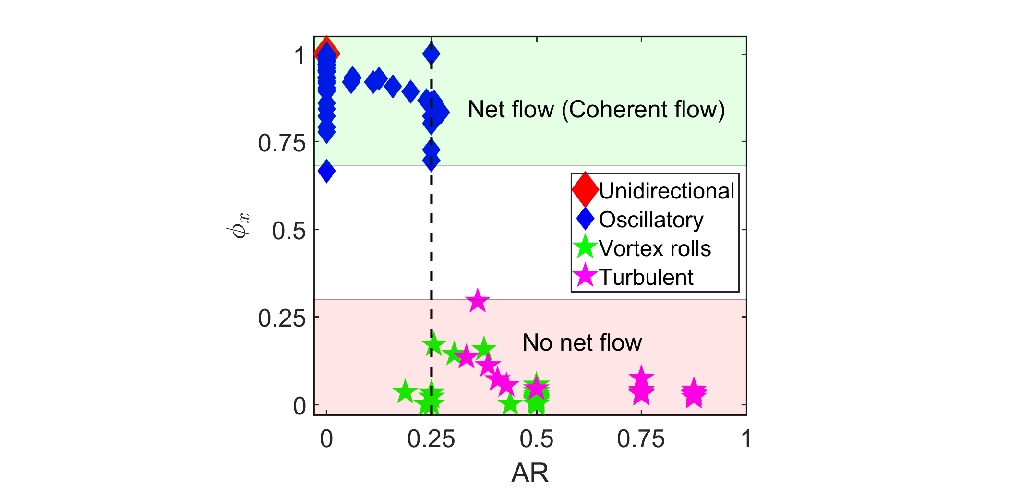} 
\caption{}
 \label{fig:Phix_align}
\end{subfigure}
\caption{ (a) Schematic of the channel used in the simulations. (b) Variation of $\phi_x$ with $\lambda$ illustrating the turbulent ($\phi_x \approx 0$) to coherent ($\phi_x \approx 1$) flow transition. The corresponding root mean square velocity, $V_{\text{rms}}$, is plotted on the secondary $y$-axis.  The dashed line indicates the $\lambda$ corresponding to the tumbling-aligning transition in a simple shear flow.  (c) Variation of $\phi_x$ with channel aspect ratio for flow aligning nematics ($\lambda = 1.0$). `$\diamond$' indicates coherent (both unidirectional and oscillatory) and `$\star$' indicates turbulent (and vortex-rolls) flows. The dashed line shows the approximate aspect ratio (AR) at which the coherent to turbulent flow transition occurs. }
\label{fig:lambda}
\end{figure*}
% Other simulation parameters: $K$ = 0.03 and $\zeta = 0.04$.
\noindent \textit{Governing equations:-} We consider a model, active nematic, incompressible fluid on a continuum scale, conserving mass and momentum~\cite{Marchetti2013,Batchelor2000}: %\ad{[Santhan, please make sure we use same convention for tensors and vectors throughout]}\\
\begin{align} 
\nabla \cdot \mathbf{u} = 0; 
% \label{eqn:mass}\\
\quad \rho \frac{D\mathbf{u}}{Dt}  = \nabla \cdot \boldsymbol{\Pi}  \label{eqn:momentum}
\end{align}
where $\rho, \mathbf{u}$ and $\boldsymbol{\Pi}$ represent the density, velocity and stress field in the fluid respectively and $\frac{D}{Dt}$ is the material derivative. 
%The flow generated by the entities of active fluid is assumed to have nematic symmetry and their
Local orientational order is described using an order parameter,
$\mathbf{Q} = {q}(3\mathbf{nn}-\mathbf{I})/2$, a second order traceless symmetric tensor field \cite{P1995}, where $q$ is the strength of the orientational order, $\mathbf{n}$ is the director field and $\mathbf{I}$ is the identity tensor. The order parameter evolves according to \cite{Giomi2013,Marenduzzo2007}
\begin{equation} 
\frac{D\mathbf{Q}}{Dt}-\mathbf{S} = \Gamma \mathbf{H}% K (\partial_k^2 Q_{ij})
\label{eqn:evolution}
\end{equation}
where
% $S_{ij}$ is the generalized 
% advection term, $S_{ij} = \left(\lambda E_{ik}+\Omega_{ik}\right)\left(Q_{kj}+\frac{\delta_{kj}}{3}\right)+\left(Q_{ik}+\frac{\delta_{ik}}{3}\right)\left(\lambda E_{kj}-\Omega_{kj}\right) -2\lambda\left(Q_{ij}+\frac{\delta_{ij}}{3}\right)\left(Q_{kl}\partial_ku_l\right)$, $\Gamma$ is the rotational
$\mathbf{S} = \left(\lambda \mathbf{E}+\boldsymbol{\Omega}\right) \cdot \left(\mathbf{Q}+{\mathbf{I}}/{3}\right)+\left(\mathbf{Q}+\mathbf{I}/{3}\right) \cdot \left(\lambda \mathbf{E}-\boldsymbol{\Omega}\right)-2\lambda\left(\mathbf{Q}+\mathbf{I}/{3}\right)\left(\mathbf{Q}:\nabla \mathbf{u}\right)$ is the generalised advection term, $\Gamma$ is the rotational diffusivity, 
$\mathbf{H} = -A_0\left(1-\frac{\gamma}{3}\right)\mathbf{Q} + A_0\gamma \left(\mathbf{Q}\cdot\mathbf{Q}-\frac{\mathbf{I}}{3}\mathbf{Q^2}\right)-A_0\gamma\mathbf{Q^2}\mathbf{Q} + K\nabla^2\mathbf{Q}$ is the molecular potential, % \sa{I have added from here} % $\mathbf{H} = -\frac{\delta F}{\delta \mathbf{Q}}+\frac{\mathbf{I}}{3}\textnormal{Tr}\left(\frac{\delta F}{\delta \mathbf{Q}}\right) = -[A_0\left(1-\frac{\gamma}{3}\right)\mathbf{Q} - A_0\gamma \left(\mathbf{Q}\cdot\mathbf{Q}-\frac{\mathbf{I}}{3}\mathbf{Q^2}\right)+A_0\gamma\mathbf{Q^2}\mathbf{Q} - K\nabla^2\mathbf{Q}]$ is the molecular potential. \sa{till here}
%$K \nabla^2 \mathbf{Q}$ is the molecular potential \cite{Chandragiri2019,Santhosh2019}.
$\mathbf{E}$ and $\boldsymbol{\Omega}$ are the symmetric and anti-symmetric parts of the velocity gradient tensor, $A_0$ sets the scale for the free energy, $\gamma$ controls the temperature and $K$ is an elastic constant determining the free energy cost of any variation in the order parameter. 
The alignment parameter, $\lambda = {9q}\lambda_1/(3q+4)$, determines the coupling between the velocity gradient and the orientational order. In the flow aligning regime, $\lambda_1 > 1$, the director field aligns at a given angle to a shear flow, while in the flow tumbling regime, $\lambda_1 < 1$ the director field rotates under shear.

The passive contributions to the stress,  $\boldsymbol{\Pi}$, are the 
Newtonian viscous stress,  $\boldsymbol{\Pi}^{\textnormal{viscous}} = 2 \mu \mathbf{E}$, and an elastic stress, 
% $\Pi_{ij}^{\textnormal{passive}} = -P\delta_{ij}+2 \lambda \left(Q_{ij}+\frac{\delta_{ij}}{3} \right) \left(Q_{kl}H_{lk} \right)
% -\lambda H_{ik} \left(Q_{kj}+\frac{\delta_{kj}}{3}\right)-\lambda \left(Q_{ik}+\frac{\delta_{ik}}{3}\right)H_{kj} -\partial_iQ_{kl}\left(K \partial_j Q_{lk} \right)
% +Q_{ik}H_{kj}-H_{ik}Q_{kj}$ 
$\boldsymbol{\Pi}^{\textnormal{passive}} = -P\mathbf{I}+2 \lambda \left(\mathbf{Q}+\mathbf{I}/{3} \right) \left(\mathbf{Q:H} \right)-\lambda \mathbf{H} \cdot \left(\mathbf{Q}+\mathbf{I}/{3}\right)-\lambda\left(\mathbf{Q}+\mathbf{I}/{3}\right) \cdot \mathbf{H} -\nabla \mathbf{Q}:\left(K \nabla \mathbf{Q} \right)+\mathbf{Q}\cdot\mathbf{H}-\mathbf{H}\cdot\mathbf{Q}$, where $\mu$ is the shear viscosity of the fluid and $P$ is the pressure. The activity of the fluid particles generates an active stress, $\boldsymbol{\Pi}^{\textnormal{active}} =-\zeta \mathbf{Q}$ \cite{Simha2002}, where $\zeta$ describes the strength of the activity. Extensile active forcing, 
$\zeta>0$, is needed to give active turbulence in an isotropic phase~\cite{Santhosh2019}.

\begin{figure*}
\begin{subfigure}{\columnwidth}
\includegraphics[width=\linewidth,trim={0 0 0 0.7cm},clip]{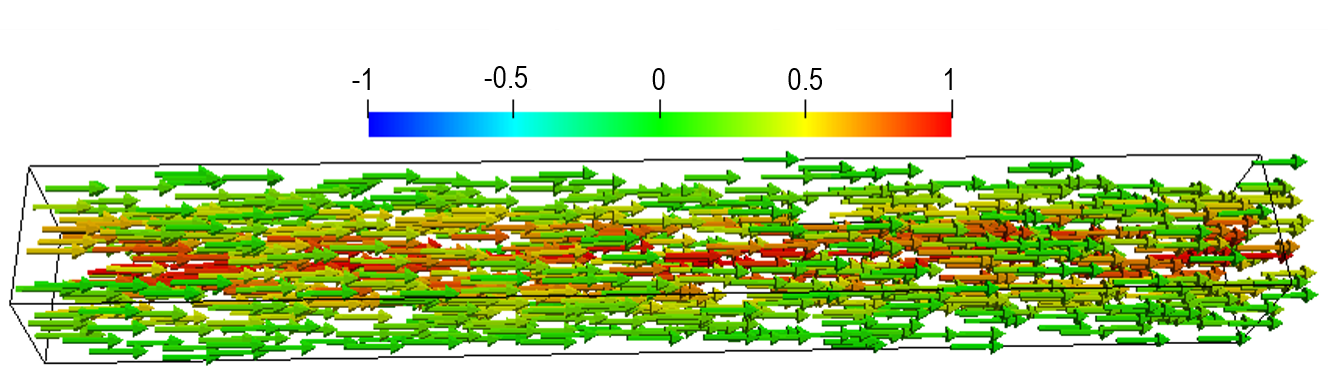}
\caption{}
\label{fig:uni_vel}
\end{subfigure}
\hspace{0.3cm}
\begin{subfigure}{\columnwidth}
\includegraphics[width=\linewidth,trim={0 0 0 0.7cm},clip]{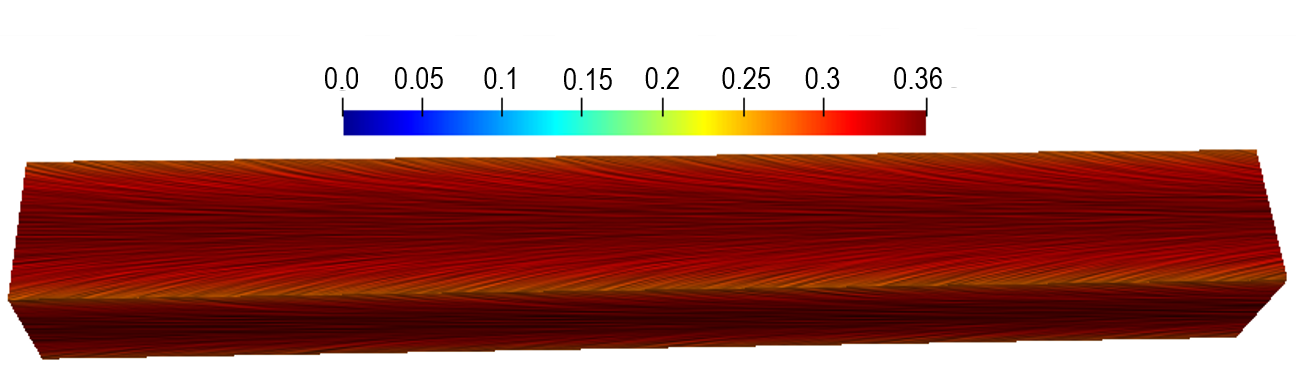}
\caption{}
\label{fig:uni_dir}
\end{subfigure}
\\
\begin{subfigure}{\columnwidth}
\includegraphics[width=\linewidth,trim={0 0 0 0},clip]{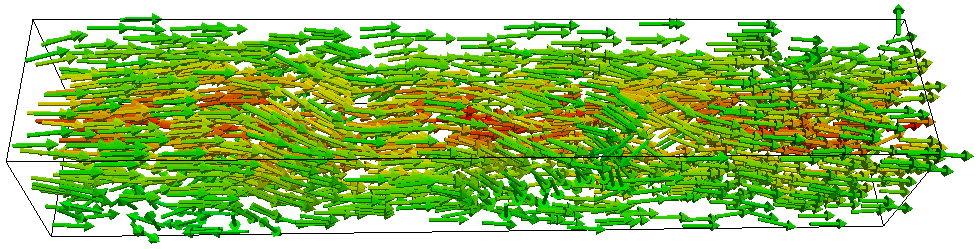}
\caption{}
\label{fig:osci_vel}
\end{subfigure}
\hspace{0.3cm}
\begin{subfigure}{\columnwidth}
\includegraphics[width=\linewidth,trim={0 0 0 0},clip]{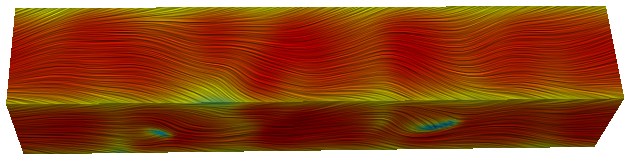}
\caption{}
\label{fig:osci_dir}
\end{subfigure}
\\
\begin{subfigure}{\columnwidth}
\includegraphics[width=\linewidth,trim={0 0 0 0},clip]{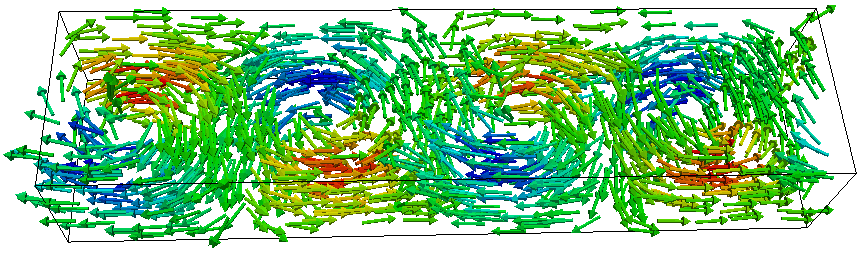}
\caption{}
\label{fig:danc_vel}
\end{subfigure}
\hspace{0.3cm}
\begin{subfigure}{\columnwidth}
\includegraphics[width=\linewidth,trim={0 0 0 0},clip]{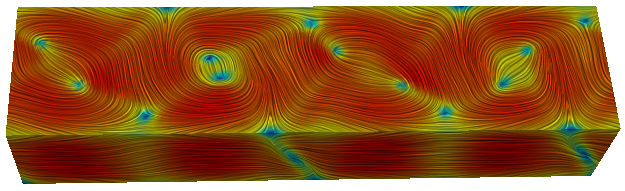}
\caption{}
\label{fig:danc_dir}
\end{subfigure}
\\
\begin{subfigure}{\columnwidth}
\includegraphics[width=\linewidth,trim={0 0 0 0},clip]{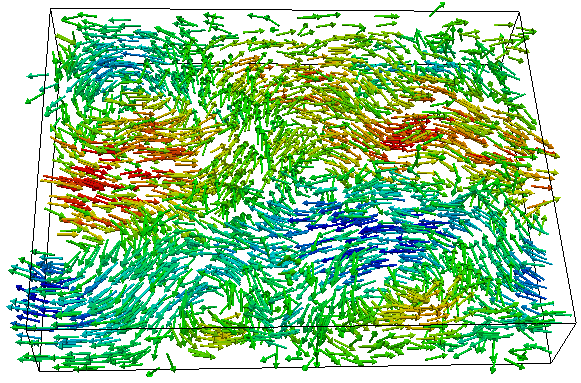}
\caption{}
\label{fig:turb_vel}
\end{subfigure}
\hspace{0.3cm}
\begin{subfigure}{\columnwidth}
\includegraphics[width=\linewidth,trim={0 0 0 0},clip]{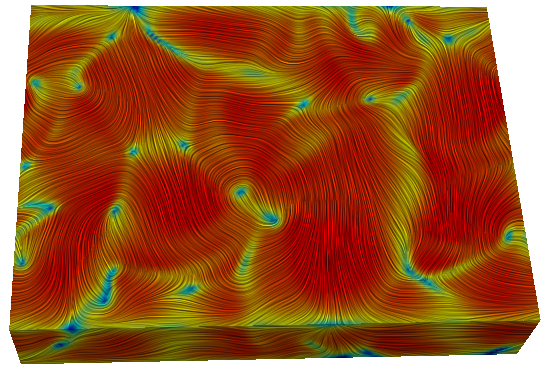}
\caption{}
\label{fig:turb_dir}
\end{subfigure}
\caption{ Active nematic flow states in a 3D channel: (a,b) unidirectional ($h = w = 16$), (c,d) oscillatory ($h = w = 24$), (e,f) lattice of vortex rolls ($h = 24, w = 32$) and (g,h) turbulent ($h = 24, w=96$) flow states. The left hand column shows the  velocity field colour coded with normalised $u_x$. The right hand column shows the director field, plotted as a line integral convolution, colour coded with the scalar order parameter. Simulation parameters are $\lambda = 1.0, K = 0.013$ and $\zeta = 0.022$.}
\label{fig:vel_profile}
\end{figure*}

\noindent\textit{Simulation details:-} The equations of motion are solved using a lattice Boltzmann method for the mass and momentum conservation equations~(\ref{eqn:momentum}) and a method of lines for the order parameter evolution equation~(\ref{eqn:evolution})~\cite{Marenduzzo2007,Doostmohammadi2017,Desplat2001}. It is not known how to map parameters in the continuum model to physical values, so we choose parameters in the range that reproduces the behaviour of 2D microtubule and motor protein mixtures~\cite{Thampi2013,Sanchez2012}, and express all quantities in lattice units. The simulation domain  (Fig.~\ref{fig:schematic}) is a channel of length $L=128$, with height, $h$ (shorter side), and width, $w$ (longer side), between 16 and 96, $\Gamma = 0.034$, $K = 0.03, \zeta = 0.04$, $\mu$ = 0.6667, $\rho = 1$ and the free energy parameters are $A_0=0.1$ and $\gamma = 2.6087$ corresponding to isotropic state of the fluid \cite{Marenduzzo2007, Chandragiri2019}.
% The LDG parameters are chosen in such a way that isotropic state is the only favourable one.} and \ju{list other parameters that don't change here}.
% Values for $K$ and $\zeta$ are given in figure captions. 
We use no slip boundary 
conditions and a no anchoring boundary condition is imposed on the orientational order parameter at the channel walls. Simulations are initialized with a stagnant fluid and a randomly oriented director field.

\noindent\textit{Results:} We begin by investigating the conditions that lead to a coherent (i.e.~a net) flow along the channel. To this end we define an order parameter
%considering a 3D channel with a square cross-section - where experiments found coherent flows were possible~\cite{Wu2017} - and examine the necessary conditions for obtaining net coherent flows along the channel. In order to quantitatively distinguish the net coherent flow state from the active turbulence state, we define a flow order parameter
%\begin{align} \label{eqn:phix}
%\phi_x= \left\langle\left|\left\langle \frac{u_x(x,y,z)}{|\mathbf{u}(x,y,z)|} \right \rangle _{x}\right|\right\rangle_{y,z},
%\end{align}
\begin{align} \label{eqn:phix}
\phi_x= \left|\left\langle \frac{u_x(x,y,z)}{|\mathbf{u}(x,y,z)|} \right \rangle _{x,y,z}\right|,
\end{align}
where $\langle \cdots \rangle_{j}$ denotes the average calculated along $j^{\textnormal{th}}$ direction. If the flow is predominantly along the channel length, $u_x >> u_y, u_z$ everywhere in the channel, then $\phi_x \rightarrow 1$, whereas for turbulent flows, where there is no net transport of fluid along the channel, $\phi_x \rightarrow 0$.

%On the other hand, for a general three dimensional flow field with no predominant flow in $x$ direction, $\phi_x \rightarrow 0$. Therefore, $\phi_x = 1$ for the net coherent flow state and $\phi_x \rightarrow 0$ for turbulent flows, where there is no net transport of fluid along the channel .

 Fig.~\ref{fig:lambda_sensitivity} shows the variation of $\phi_x$ with the flow-aligning parameter $\lambda$  for a square channel with $w = h = 16$. It is evident that increasing $\lambda$ results in a sharp crossover from no net flow in the flow-tumbling regime to coherent flow when the fluid becomes flow-aligning. We find that this crossover is robust to changing activity, $\zeta$, orientational elasticity, $K$, and the size 
%  \sout{and shape} 
 of the channel.
%  \sa{[being checked]}% \ju{check}}.
 %$w$?, mention explicitly that this result is for square channels, although we believe it is always true?} 
 %indicating that in a channel with square cross-section, net coherent flows are only possible for the particles with flow-aligning behaviour.\\ 
 
 We next restrict ourselves to the flow-aligning regime and, guided by the experiments~\cite{Wu2017}, change the aspect ratio of the channel cross-section. To 
%Interestingly, upon changing the channel cross-section the flow behvaiour begins to alter and the coherent flows diminish. To 
quantify the aspect ratio we define AR = ${(w-h)}/w$.
% where $\text{max}(h,w)$ is the larger of $h$ and $w$.
For square channels AR$ = 0$, whereas if 
$h$ and $w$ differ significantly,  AR $\rightarrow 1$. 

Fig.~\ref{fig:Phix_align} summarises the geometry dependence of the coherent to turbulent flow transition by measuring the flow order parameter, $\phi_x$, 
as a function of the aspect ratio of the channel.  Interestingly, there is a sharp  transition between coherent flow states and active turbulence as the aspect ratio increases beyond $\sim 0.25$.

%The net flow states are only observed when $AR \lesssim 0.25$, with a sharp transition between the two regimes.

%, separately for flow aligning (Fig.~\ref{fig:Phix_align}) and flow tumbling (Fig.~\ref{fig:Phix_tumble}) fluid. The net coherent flow states are only observed for flow aligning active nematic in symmetric channels, i.e, when $AR \lesssim 0.25$. In asymmetric channels, i.e, when $AR \gtrsim 0.25$, no net coherent flow is observed for the same active fluid. Therefore, the aspect ratio $AR$ clearly distinguishes net flow states from no net flow states. This transition from a net flow to a no net flow state as a function of aspect ratio is very sharp, $\phi_x$ changes from $\approx 0.7$ to $\approx 0.2$ without the existence of any intermediate values, reminiscent of a first order phase transition \ad{[I am not sure if we can claim this]}. On the other hand a flow tumbling active nematic only exists in a no net flow state irrespective of the aspect ratio of the channel, and no transition to a net coherent flow is observed (Fig.~\ref{fig:Phix_tumble}). Taken together these results demonstrate that the material property of flow alignment and a geometry with a symmetric cross section are two prerequisites to obtain a net transport of active isotropic fluid in a channel.

To understand this behaviour we use the simulations  to more closely examine the details of the flow structure close to the transition from net flow to the active turbulent state. Near the transition point, AR $\approx 0.25$, neither purely laminar flow nor fully-developed turbulent flow are observed. Instead, the active fluid can demonstrate oscillatory flow or a lattice of vortex-rolls. The different flow states observed as the channel size
% (or, equivalently, the activity \sa{being checked})
is increased are shown in Fig.~\ref{fig:vel_profile}.

For very narrow channels the active instability is suppressed and there is no flow. An increase in the channel size first leads to active flows that are
completely unidirectional, with a velocity vector that only has a component along the channel length, $\mathbf{u} = u_x (y,z) \hat{\mathbf{x}}$ (Fig.~\ref{fig:uni_vel}). %Thus, the velocity field is one-dimensional as shown in Fig.~\ref{fig:uni_vel}, and it is similar to a pressure driven laminar flow of a fluid in a rectangular channel \cite{Leal2007}. 
For slightly wider channels, the flows can develop an oscillatory component and $u_y \neq u_z \neq 0$ (Fig.~\ref{fig:osci_vel}). However, the velocity along the channel remains the dominant component of the velocity field with its maximum value near the center line of the channel. Since  $u_x$ is dominant in both unidirectional and oscillatory flows, these both manifest as coherent states which result in a net fluid transport through the channel.

Further increases in the size of the channel  can, however, lead to flow configurations where the net flow is absent. The vortex-roll state is characterized by 
 three-dimensional,  counter-rotating vortices located on a one-dimensional lattice along the length of the channel, as shown in Fig. \ref{fig:danc_vel}. The axes of the vortices invariably lie along the smallest channel dimension ($y$ direction). On any cross-section perpendicular to the $y$ direction, the 3D flow appears similar to the dancing flows reported earlier in 2D simulations~\cite{Shendruk2017,Chandragiri2019} and 2D confined microtubule and motor protein mixtures~\cite{Hardouin2019}. 
 Regions with $q=0$ appear in
 the corresponding director field  indicating the presence of disclination lines (Fig.~\ref{fig:danc_dir}). These structures are dynamic and may form either in the bulk or near the walls. Finally, in large channels, we recover active turbulence, characterised by a spatio-temporally evolving director field (Figs.~\ref{fig:turb_vel},~\ref{fig:turb_dir}) which results in contortion of disclination lines and their irregular spatial arrangement~\cite{Shendruk18,Binysh20,Duclos20}.
 
\begin{figure}
\begin{subfigure}{0.47\linewidth}
    \includegraphics[width=1.6\linewidth,trim={9cm 0 -2cm 0},clip]{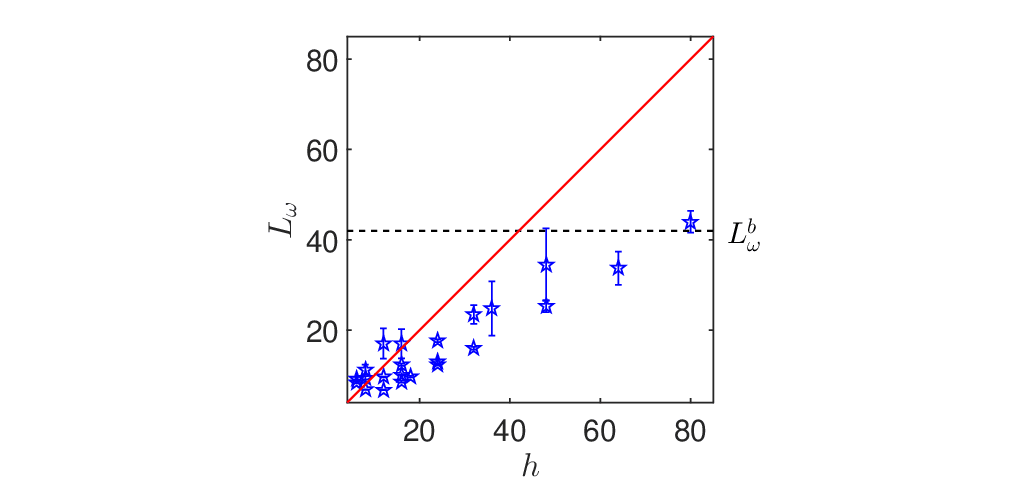} 
\caption{}
 \label{fig:vorticity_trend}
\end{subfigure}
\begin{subfigure}{0.47\linewidth}
    \includegraphics[width=1.6\linewidth,trim={9cm 0 -2cm 0},clip]{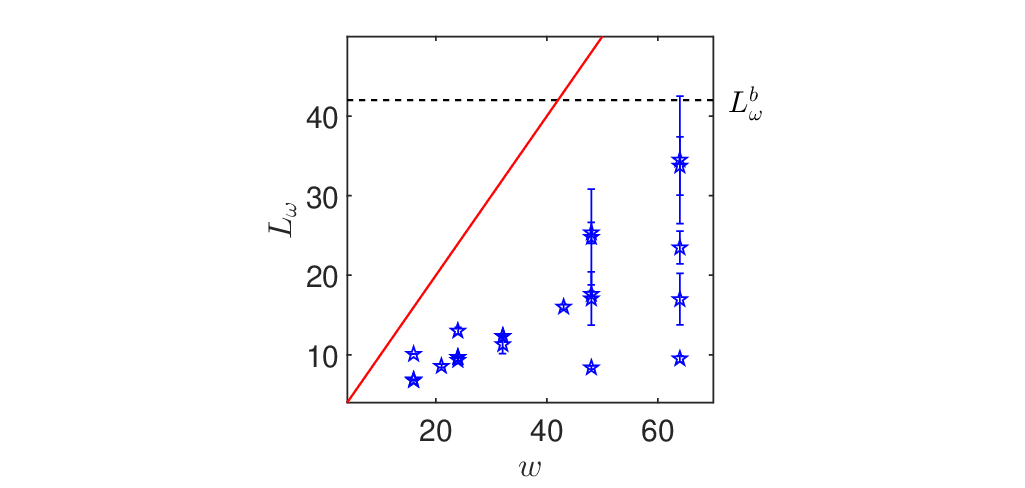}
\caption{}
\label{fig:vorticity_length}
\end{subfigure}
\caption{Variation of vorticity correlation length against channel dimensions: (a) $L_\omega$ vs $h$ and (b) $L_\omega$ vs $w$. If $\boldsymbol{\omega} = \nabla \times \mathbf{u}$ then $L_\omega$ is determined as the distance $r$ at which the correlation function $\langle \boldsymbol{\omega}(r)\cdot \boldsymbol{\omega}(0)\rangle$ calculated in the $xz$ plane at $y = h/2$ decays to zero. $L_\omega^{b}$ is the vorticity correlation length in the bulk calculated from bulk simulations. The solid line represents the line $x = y$
% is of slope unity and 
and the dashed line is drawn at $L_\omega = L_\omega^b$.}
\label{fig:vorticity}
\end{figure}

\noindent\textit{Mechanism:-} We can now explain the disappearance of net flow as the aspect ratios of the channels increase.
% \sa{\sout{Our argument is based on the hydrodynamic screening caused by the channel walls and the emergence of vortex rolls near the transition aspect ratio.}}
Firstly we note that, as a result of hydrodynamic screening, the vorticity correlation length in the channel, $L_\omega$, is set by its smallest dimension, $h$. Evidence for this is presented in Fig.~\ref{fig:vorticity_trend} where we show that the vorticity correlation length $L_\omega$, (measured in the $xz$ plane at $y = h/2$) tracks $h$ until the channel becomes too wide to screen the flows and the correlation length crosses over to its bulk value $L_\omega^b$. For comparison, Fig.~\ref{fig:vorticity_length} shows that there is no correlation between
$L_\omega$ and the larger dimension of the channel cross section $w$.
%the vorticity correlation length and max($w,h$) .
However, the structure of the flow is determined by $w$.
%the maximum value of $w$ and $h$. 
%If max($w,h$) $\lesssim$ $\omega$,
If $w$ $\lesssim$ $L_\omega$, vortices cannot form and there is a net flow. If, however, 
%max($w,h$) $\gtrsim$ $\omega$
$w$ $\gtrsim$ $L_\omega$
the larger dimension of the channel cross section becomes available for the flow streamlines to turn, form vortices, and destroy the coherent net flow along the channel.  This also explains why the vortex rolls are always ordered along the channel length with their axis along the shortest dimension of the channel. No analytical estimate for the exact value of the coefficient in these inequalities is available. However numerical simulations of 2D confined active nematics~\cite{Shendruk2017} suggest that $w \sim 1.4\times h$ when vortex rolls first form. This corresponds to $AR \sim 0.28$. Note that the argument breaks down when both $w$ and $h$ are larger than the bulk correlation length, $L_\omega^b$: active turbulence will then destroy any net flow regardless of the aspect ratio.
 
% at the transition. For the net coherent flow to form in a 3D channel with the cross section of $h,w$ dimensions, we should have $h>h^*$ and $w<w^*$ or $w/h<w^*/h^*\sim 1.4$ corresponding to $AR \gtrsim 0.28$.

 %First, let us consider the $x-y$ plane - the plane formed by the channel length and the smallest dimension of the channel cross-section. For any given activity and orientational elasticity, the spontaneous flow transition~\cite{Voituriez2005} occurs when the smallest dimension of the channel ($h$) becomes larger than the threshold $h^*=\left(8\eta K\pi^2/\zeta(1-\lambda)\Gamma\right)^{1/2}$ for a transition to net flow in a 2D confined active nematics~\cite{Voituriez2005,Giomi2012}. Now, consider the $x-z$ plane - the plane formed by the channel length and the biggest dimension of the channel cross-section. 
 
%No analytical estimate for the formation of such vortices is available, however numerical simulations of 2D confined active nematics~\cite{Shendruk2017} suggest that $w^* \sim 1.4\times h^*$. For the net coherent flow to form in a 3D channel with the cross section of $h,w$ dimensions, we should have $h>h^*$ and $w<w^*$ or $w/h<w^*/h^*\sim 1.4$ corresponding to $AR \gtrsim 0.28$.

 %%%%%%%%%%%%%%%%%%%%%%%%%%%%%
%%%%%%%%%%%%%%%%%%%%%%%%%%%%%

\noindent\textit{Discussion:-} Our results demonstrate that, in flow-aligning, 3D, active microfluidics, the emergence of coherent net flow states is controlled by the aspect ratio of the channel cross section: coherent flows are possible in channels with cross sections that are close to isotropic, but are destroyed in channels with larger aspect ratios. This can be explained by whether or not vortex rolls have room to form across the larger cross-section dimension, thus destroying coherent flow. The length scale of the vortices is set by hydrodynamic screening controlled by the smaller dimension. 

%Therefore, compared to the 2D confined active flows, the channel aspect ratio appears as an additional control parameter for transition between flow states in 3D. 
The aspect-ratio dependence agrees with recent experiments on microtubule-motor protein mixtures in microfluidic channels~\cite{Wu2017}. 
However,
% \sa{\sout{it is instructive to note that}} 
to more closely compare the results with the experimental system, the model in its current form requires a number of adjustments. First, here we assume a uniform density  throughout the channel, while the experiments clearly indicate that layers of concentrated aligned microtubules can build up  at the confining boundaries. %This can be accommodated within the model by introducing additional order parameter that allows for concentration variation~\cite{Giomi2012,Thampi2015}. 
Indeed, the experimental results were attributed to such surface ordering, but we show here that  the aspect ratio induced flow transition does not require wetting of the channel walls by a nematic layer.
Second, it is possible that free-slip velocity boundary conditions and weak-planar anchoring of the orientation might result in a more accurate representation of experiments as the microtubules appear to weakly align and slide freely at the boundaries~\cite{Hardouin2019,Opathalage2019}.  We also show that the transition relies upon the active fluid being flow-aligning, thus identifying the microtubule - motor protein mixtures as belonging to this class of nematic fluids.

\bibliography{references3.bib}
\end{document}